\pdfoutput=1
\documentclass[11pt]{article}
\usepackage[preprint]{acl}
\usepackage{times}
\usepackage{enumitem}
\usepackage{tikz}
\usetikzlibrary{shapes, arrows.meta, positioning}
\usepackage{booktabs}
\usepackage{multirow}
\usepackage{latexsym}
\usepackage{amsmath}  
\usepackage{amssymb}   
\usepackage{tikz} 
\usepackage{hyperref}
\usepackage{xcolor}
\usepackage[T1]{fontenc}
\usepackage[utf8]{inputenc}
\usepackage{microtype}
\usepackage{amsmath,amssymb}       
\usepackage{algorithm}            
\usepackage{algorithmicx}          
\usepackage{algpseudocode}         
\usepackage{inconsolata}
\usepackage{graphicx}
\usepackage{subcaption, url,booktabs}
\usepackage{xcolor,graphicx,colortbl}
\usepackage{booktabs}
\usepackage{multirow}

\definecolor{emerald5}{RGB}{236,247,238}   
\definecolor{emerald10}{RGB}{200,230,210}  
\definecolor{emerald15}{RGB}{164,214,185}

\title{\textit{Curved Worlds, Clear Boundaries:} Generalizing Speech Deepfake Detection using Hyperbolic and Spherical Geometry Spaces}

\author{
 \textbf{Farhan Sheth\textsuperscript{1}\thanks{Equal Contribution as a first authors}},
 \textbf{Girish\textsuperscript{2}\footnotemark[1]}
 \textbf{Mohd Mujtaba Akhtar\textsuperscript{3}\footnotemark[1]},
 \textbf{Muskaan Singh\textsuperscript{4}} 
\\
 \textsuperscript{1}Manipal University Jaipur, India,
 \textsuperscript{2}UPES, India,
 \textsuperscript{3}V.B.S.P.U, India,
 \textsuperscript{4}Ulster University, UK,
\\
 \small{
    \textbf{Correspondence:} \href{mailto:email@domain}{m.singh@ulster.uk.in}
 }
}

\begin{document}
\maketitle

\begin{abstract}
In this work, we address the challenge of generalizable audio deepfake detection (ADD) across diverse speech synthesis paradigms—including conventional text-to-speech (TTS) systems and modern diffusion or flow-matching (FM) based generators. Prior work has mostly targeted individual synthesis families and often fails to generalize across paradigms due to overfitting to generation-specific artifacts. We hypothesize that synthetic speech, irrespective of its generative origin, leaves behind shared structural distortions in the embedding space that can be aligned through geometry-aware modeling. To this end, we propose \textbf{\texttt{RHYME}}, a unified detection framework that fuses utterance-level embeddings from diverse pretrained speech encoders using non-Euclidean projections. \textbf{\texttt{RHYME}} maps representations into hyperbolic and spherical manifolds—where hyperbolic geometry excels at modeling hierarchical generator families, and spherical projections capture angular, energy-invariant cues such as periodic vocoder artifacts. The fused representation is obtained via Riemannian barycentric averaging, enabling synthesis-invariant alignment. \textbf{\texttt{RHYME}} outperforms individual PTMs and homogeneous fusion baselines, achieving top performance and setting new state-of-the-art in cross-paradigm ADD.
\end{abstract}

\section{Introduction}

The field of speech synthesis has undergone a paradigm shift over the past few years with the emergence of neural text-to-speech (TTS) and diffusion-based models, delivering human-level audio realism that is increasingly difficult to distinguish from genuine speech \citet{yi2023audio}. While these technological advances have enabled beneficial applications in accessibility and entertainment, they have simultaneously paved the way for malicious applications. Recent high-profile cases—such as voice clones used in emotional blackmail and scam calls—highlight the urgent need for reliable detection mechanisms, since human listeners are no longer able to consistently identify deepfake audio \citet{barrington2025people}. In response, audio deepfake detection (ADD) has emerged as a crucial research domain within speech security and digital forensics. Despite notable advances in audio deepfake detection, most existing systems are trained on speech generated by conventional TTS or vocoder-based models and demonstrate strong performance under seen conditions. However, these systems often exhibit degraded performance when exposed to synthetic speech generated using unseen paradigms or under distributional shift. \citet{chen2020generalization} identified this issue early on, showing that detectors often fail even within the same benchmark when exposed to unfamiliar spoofing techniques. Extending this analysis, \cite{kulkarni2024exploring} evaluated a range of self-supervised models across datasets and synthesis types, reporting consistent performance drops in cross-domain settings. These findings highlight a fundamental limitation in existing ADD pipelines and point toward the need for detection strategies that are resilient to synthesis diversity and adaptable to new generation mechanisms. Although neural speech generators differ widely in architecture and training objectives, we hypothesize that synthetic speech—regardless of whether it is produced by TTS, vocoder, or diffusion-based models—shares certain latent artifacts that distinguish it from natural speech. These artifacts may not be easily perceptible in the waveform or spectrogram, but they can manifest as distortions in the underlying feature space. By leveraging diverse pre-trained speech encoders, it is possible to capture complementary representations that highlight different aspects of these synthesis-induced irregularities. The central challenge lies in aligning these heterogeneous embeddings in a way that preserves discriminative cues while minimizing overfitting to a specific synthesis family. This motivates our exploration of a shared, synthesis-invariant representation space for generalized detection; in this work, we emphasize acoustic evidence, analyzing signal-level artifacts while remaining independent of language or text. \par
To this end, we propose \textbf{\texttt{RHYME:}} \textbf{R}iemannian fusion of \textbf{HY}perbolic and sph\textbf{E}rical embeddings a unified and geometry-aware audio deepfake detection framework. RHYME fuses utterance-level embeddings from multiple frozen speech PTMs and projects them into two complementary non-Euclidean manifolds: hyperbolic space (to capture hierarchical generative traces) and spherical space (to model periodic and spectral anomalies). These projections are then fused using Riemannian barycentric averaging in the Poincaré ball, forming a synthesis-agnostic representation space.
Each of these geometric components serves a distinct purpose—hyperbolic projection models generator hierarchies, spherical projection highlights periodic artifacts, and barycentric fusion unifies them while preserving their respective curvatures. Hyperbolic curvature compactly encodes tree-like generator lineages (e.g., vocoder $\rightarrow$ autoregressive TTS (AR-TTS) $\rightarrow$ diffusion/flow) with low distortion, placing related synthesizers along nearby geodesics while genuine speech occupies distinct regions. Consequently, the hyperbolic branch separates synthetic from real speech by tracing these geodesic hierarchies (model families/versions). As newer diffusion or flow-based methods still follow a structured progression of architectural improvements, their embeddings naturally fall on new branches of this learned hyperbolic hierarchy. In parallel, the spherical branch captures angular, energy-invariant cues associated with periodicity artifacts, such as those introduced by vocoders or diffusion sampling (See Section \ref{ModelRHYME} for details). Even though waveform-level characteristics vary, these models still imprint subtle periodic distortions, which manifest as consistent angular deviations on the hypersphere. By fusing these complementary cues via Riemannian averaging, \textbf{\texttt{RHYME}} ensures that unseen synthetic speech cannot simultaneously satisfy both the hierarchical and periodic constraints of genuine speech—resulting in robust generalization across synthesis families. \textit{We hypothesize that this geometry-aware fusion encourages synthesis-invariant alignment while preserving discriminative cues necessary for deepfake detection.} We validate \textbf{\texttt{RHYME}} through comprehensive experiments on two widely-used benchmark datasets—ASVspoof and DFADD—which include speech synthesized using TTS, vocoder, and diffusion-based models. Our evaluation covers zero-shot, cross-corpus, and unseen-generator scenarios to rigorously test generalization. Across all settings, \textbf{\texttt{RHYME}} consistently delivers strong performance, even when faced with synthetic speech from generators it has never seen during training. Unlike previous methods that tend to overfit to specific synthesis families, our geometry-aware fusion strategy enables \textbf{\texttt{RHYME}} to learn shared structural cues across different types of fake speech. This shows that RHYME handles both familiar and previously unseen generators well, making it a strong candidate for practical deployment in deepfake detection systems. \newline

\noindent \textbf{Main Contributions:} To this end, in this paper, we present the following key contributions:
\begin{itemize}
    \item We introduce \textbf{RHYME}, a unified and geometry-aware framework for generalizable audio deepfake detection across TTS, vocoder, and diffusion-based speech generators.
    \vspace{-0.2cm}
    \item We leverage non-Euclidean projections, where hyperbolic geometry captures hierarchical relationships among generators, and spherical space models angular, periodic artifacts commonly introduced by vocoding and diffusion processes.
    \vspace{-0.2cm}
    \item We propose a novel Riemannian fusion strategy that uses barycentric averaging in the Poincaré ball to align heterogeneous pretrained embeddings into a shared, synthesis-invariant representation space.
    \vspace{-0.2cm}
    \item We demonstrate that \textbf{RHYME} consistently outperforms individual PTMs and homogeneous fusion baselines, achieving state-of-the-art performance under cross-corpus and unseen-generator conditions.
    \vspace{-0.2cm}
    \item We extend the DFADD benchmark by adding two new diffusion-based generators, providing a more rigorous testbed for evaluating out-of-distribution generalization in future work.
\end{itemize}
\vspace{-0.2cm}
\noindent \textit{The source code and models can be accessible at \footnote{\url{https://github.com/Helixometry/RHYME-IJCNLP-AACL.git}.}}

\section{Related Work}
The field of audio deepfake detection (ADD) originated from the vulnerabilities identified in automatic speaker verification (ASV) systems. Early countermeasures utilized handcrafted spectral features such as constant-Q cepstral coefficients (CQCC) and mel-frequency cepstral coefficients (MFCC), often paired with Gaussian mixture models or support vector machines. Early research in this area primarily leveraged handcrafted spectral features such as constant-Q cepstral coefficients (CQCC) and linear frequency cepstral coefficients (LFCC), often in combination with traditional machine learning classifiers for spoofing detection tasks \cite{todisco2017constant,kinnunen2017asvspoof}. The advent of the ASVspoof challenge series \cite{todisco2019asvspoof,wang2020asvspoof} propelled the field toward deep learning solutions, with convolutional and recurrent neural networks trained on spectrograms or raw waveforms becoming standard for text-to-speech (TTS) and voice conversion (VC) based deepfake detection \cite{cai2023face}. However, most of these methods are tailored to specific synthesis techniques and struggle to generalize to audio produced by newer paradigms such as diffusion and flow-matching models. The emergence of self-supervised learning (SSL) has significantly advanced the capabilities of audio deepfake detection. Models such as wav2vec 2.0 \cite{baevski2020wav2vec}, HuBERT \cite{hsu2021hubert} and WavLM \cite{chen2022wavlm} have demonstrated robust performance across various paralinguistic and spoofing benchmarks, including ASVspoof and ADD tasks. These models have since been adapted to a range of spoofing detection tasks. \cite{tak2022automatic} investigated the efficacy of wav2vec 2.0 in detecting synthetic audio under various ASVspoof 2021 conditions, highlighting the potential of SSL models in practical ADD scenarios. Building on this line of work, \cite{guo2024audio} proposed a multi-level fusion framework based on WavLM, achieving competitive performance across multiple spoofing benchmarks including ASVspoof 2019 and 2021. Building on this body of work, \cite{kheir2025comprehensive} conducted a detailed layer-wise study of multiple SSL models and found that lower transformer layers consistently offer the most discriminative features for deepfake detection across languages and tasks. These insights reinforce the potential of lightweight, generalizable detection pipelines. Nevertheless, most existing methods remain limited to single-model fine-tuning, restricting their adaptability to unseen synthesis techniques—an issue our work seeks to address through a fusion-driven, synthesis-agnostic approach. Despite progress in ADD, a key limitation persists in generalizing across diverse synthesis paradigms. Most existing detectors are trained primarily on speech generated by conventional text-to-speech (TTS) systems and struggle to maintain performance when exposed to speech produced using newer generative methods such as diffusion or flow-matching (FM) models. To address this shift, recent studies have proposed fusion-based frameworks that leverage multimodal foundation models \cite{chetia2025towards} and paralinguistic speech encoders \cite{akhtar2025source} to improve robustness in source attribution tasks. These approaches have shown the benefit of combining heterogeneous pre-trained models to capture generator-specific characteristics. Motivated by this direction, our work targets the detection problem under similar distribution shifts, specifically focusing on bridging generalization gaps across synthesis families. We aim to learn synthesis-invariant embeddings by fusing multiple speech encoders and projecting their representations into a hyperspherical space. This allows our model to detect synthetic speech reliably across both known and unseen generation mechanisms, including diffusion-based generators.

\begin{figure*}[!h]
    \centering
    \includegraphics[width=0.971\linewidth]{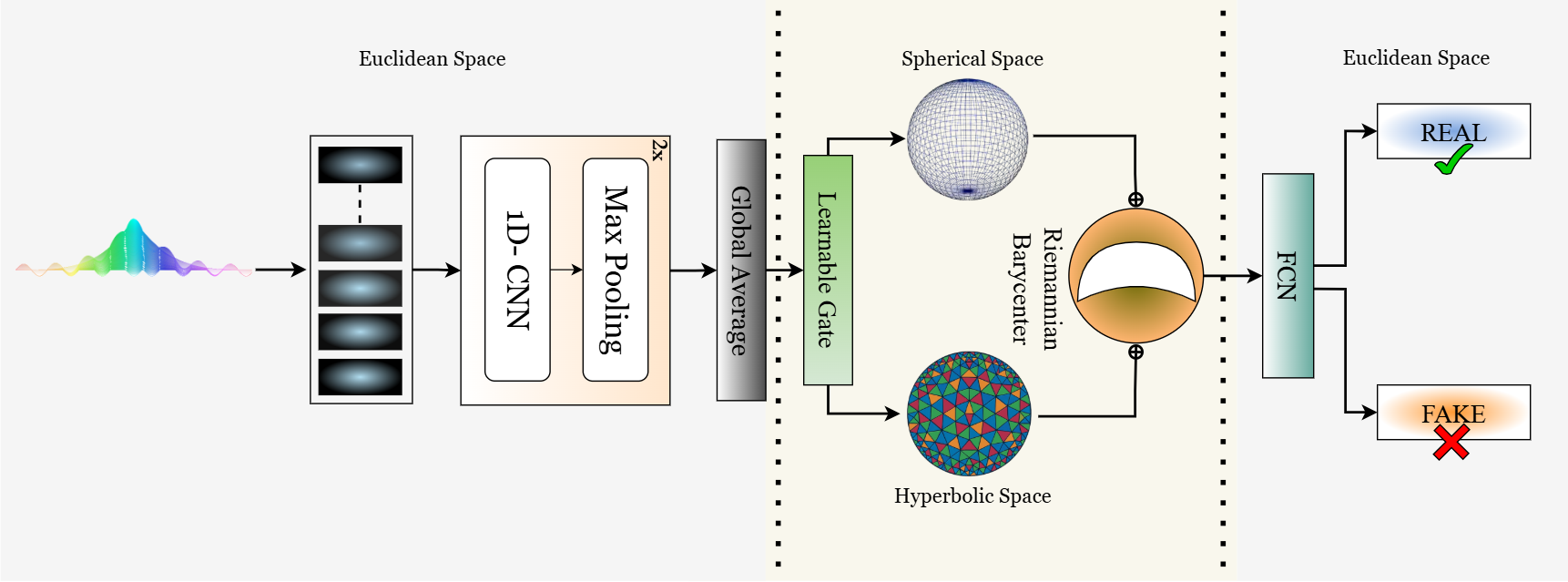}
    \caption{Working flow of the proposed \textbf{\texttt{RHYME}} framework.}
    \label{fig:1}
\end{figure*}

\section{Pre-trained Models}
In this section, we describe the speech representation models considered in our study. \newline
\noindent USAD\footnote{\url{https://huggingface.co/collections/MIT-SLS/usad-models-68491d4c7d0978b85d0c4299}} \cite{chang2025usad} is a universal audio representation model trained using multi-teacher distillation across speech, music, and sound domains, and we adopt its Base variant (94M parameters). PaSST\footnote{\url{https://github.com/kkoutini/PaSST}} \cite{koutini2021efficient} adapts vision transformers for spectrograms with patch-masking for regularization; we use the PaSST-S model (87M). Whisper\footnote{\url{https://huggingface.co/openai/whisper-base}} \cite{radford2023robust}, a multilingual ASR model trained on 680k hours of weakly supervised audio-text data, is used in its Base variant (74M), and we extract encoder embeddings. x-vector\footnote{\url{https://huggingface.co/speechbrain/spkrec-xvect-voxceleb}} \cite{8461375}, a 4.2M parameter TDNN trained for speaker recognition, is included as a lightweight yet strong baseline and has proven effective for deepfake-related tasks. WavLM\footnote{\url{https://huggingface.co/microsoft/wavlm-base}} \cite{chen2022wavlm} builds on Wav2Vec 2.0 with masked prediction and mixture modeling and we use its Base version (94M) pretrained on 94K hours of audio. HuBERT\footnote{\url{https://huggingface.co/facebook/hubert-base-ls960}} \cite{hsu2021hubert} is included for its phonetic and prosodic modeling ability, and we adopt the Base version (95M) trained on 960h of LibriSpeech. Lastly, Wav2Vec 2.0\footnote{\url{https://huggingface.co/facebook/wav2vec2-base}} \cite{baevski2020wav2vec} is a contrastive SSL model for speech representation learning; we use the Base variant (94M), which provides strong acoustic modeling directly from raw waveforms and serves as a solid backbone for capturing subtle synthetic patterns in speech. \newline
\noindent We extract representations from the last hidden state of the frozen PTMs using average pooling. The resulting representation dimensions are: 768 for USAD and PaSST; 768 for WavLM, Wav2Vec2, and HuBERT; 512 for x-vector; and 768 for Whisper (Base). All audio samples are resampled to 16kHz before feeding into the PTMs.

\section{Modeling Pipeline}

\subsection{Proposed Framework: \textbf{\texttt{RHYME}}}
\label{ModelRHYME}
We propose, \textbf{\texttt{RHYME}} for generalizable audio deepfake detection across diverse synthesis models. The architecture is presented in Figure~\ref{fig:1}. \newline
\noindent We begin by encoding the input waveform. Let \(x(t)\) be the raw audio waveform. We first encode \(x(t)\) using a frozen self-supervised speech-foundation model \(g\):
\begin{equation}
s(\tau) = g\bigl(x(t)\bigr)\;\in\;\mathbb{R}^D,\quad \tau=1,\dots,T.
\end{equation}

\noindent We then pass the frame-level embeddings through a 1D convolutional encoder \(\Phi_{\mathrm{1D}}\) to model local temporal patterns:
\begin{equation}
H(p,\tau') = \Phi_{\mathrm{1D}}\bigl(s(\tau)\bigr)\;\in\;\mathbb{R}^{C\times T'}.
\end{equation}

\noindent We apply global average pooling over the time axis to obtain an utterance-level representation:
\begin{equation}
u = \frac{1}{T'}\sum_{\tau'=1}^{T'} H(:,\tau')\;\in\;\mathbb{R}^d.
\end{equation}

\noindent We learn a gating scalar \(\alpha\) using a sigmoid activation to adaptively control the separation into geometric branches:
\begin{equation}
\alpha = \sigma\bigl(w_g^\top u + b_g\bigr)\;\in\;(0,1)
\end{equation}

\noindent We then split \(u\) into two components:
\[
u_h = \alpha\,u,\qquad
u_s = (1-\alpha)\,u.
\]

\noindent These gated components are subsequently projected into distinct non-Euclidean manifolds for geometry-aware fusion.

\paragraph{Hyperbolic branch}  
We project the gated vector \(u_h\) into the Poincaré ball \(B^d_c\) using the exponential map with curvature \(c>0\):
\begin{equation}
x_h = \exp^{c}(u_h) = \tanh\!\bigl(\sqrt{c}\,\lVert u_h\rVert\bigr)\;
\frac{u_h}{\sqrt{c}\,\lVert u_h\rVert} \;\in\;B^d_c.
\end{equation}
\paragraph{Spherical branch}  
We normalize the component \(u_s\) onto the unit sphere and map it into the same Poincaré ball via stereographic projection:
\begin{align}
x_s &= \frac{u_s}{\lVert u_s\rVert}\;\in\;S^{d-1},\\
y_s &= \sigma^{-1}(x_s) = \frac{x_s}{1 + \sqrt{1 - \lVert x_s\rVert^2}} \;\in\;B^d_c.
\end{align}

\paragraph{Riemannian Barycentric Fusion}

We compute the Riemannian barycenter of the projected embeddings \(x_h\) and \(y_s\) to perform non-Euclidean fusion in the hyperbolic space:
\begin{equation}
z^* = \exp^{c}\!\Bigl(\alpha\,\log^{c}(x_h) + (1-\alpha)\,\log^{c}(y_s)\Bigr) \;\in\;B^d_c,
\end{equation}
where \(\log^{c}(\cdot)\) denotes the inverse exponential map at the origin.

\paragraph{Euclidean Representation and Classification}

We project the fused representation \(z^*\) back to Euclidean space using the logarithmic map:
\begin{equation}
r = \log^{c}(z^*)\;\in\;\mathbb{R}^d.
\end{equation}

\noindent We pass this vector through a lightweight linear classifier to obtain prediction logits:
\begin{align}
\ell &= W_{\mathrm{cls}}\,r + b_{\mathrm{cls}},\\
\hat y &= \text{softmax}(\ell),
\end{align}

\noindent We train the entire model—including \(\Phi_{\mathrm{1D}}\), \(w_g, b_g\), curvature \(c\), and classifier weights—end-to-end using cross-entropy loss:
\begin{equation}
\mathcal{L} = -\sum_{k=0}^K y_k \,\log \hat y_k.
\end{equation}
\noindent Ensuring that $\lVert z^* \rVert < 1$ maintains numerical stability within the hyperspherical fusion space. A fully connected classifier with a dense layer is attached to the Euclidean representation $r$, followed by a softmax output that predicts the probability of the input being real or synthetic. \textbf{\texttt{RHYME}} has a parameter footprint ranging from 8 to 14 million, depending on the combination and dimensionality of the pretrained speech encoders used.

\section{Experiments}

\subsection{Benchmark Dataset}
\label{corpus}
Our study is grounded in experiments on two datasets: \newline
\noindent\textbf{DFADD} \cite{du2024dfadd}\footnote{\url{https://github.com/isjwdu/DFADD}}: It introduces a new generation of highly realistic spoofed audio samples generated using diffusion and flow-matching (FM) text-to-speech models. It includes over 163,500 synthetic samples paired with 44,455 bonafide utterances from the VCTK corpus, covering 109 speakers. The spoofed audio is synthesized using Diffusion-based and Flow-matching-based speech generation models.  \newline
\noindent\textbf{ASVSpoof 2019 (ASV)} \cite{wang2020asvspoof}\footnote{\url{https://datashare.ed.ac.uk/handle/10283/3336}}: The dataset is a widely adopted benchmark for evaluating spoof detection systems under both logical access (LA) and physical access (PA) scenarios. In our work, we utilize the LA subset, which contains speech spoofed using traditional TTS and voice conversion (VC) methods—primarily waveform concatenation, parametric synthesis, and neural vocoders. The subset consists of bonafide utterances from 20 speakers sourced from the VCTK corpus Spoofed audio generated using 17 different TTS and VC systems, including some previously unseen systems in the evaluation set. We adopt the original train/test split of the LA subset as specified in the paper. \newline
\noindent\textbf{Dataset Usage Protocol}:
In our study, we use these datasets to simulate cross-paradigm generalization. Specifically, we train on traditional TTS data from ASV and evaluate on DFADD to measure forward generalization from older to newer synthesis methods. Conversely, we also train on DFADD and evaluate on ASV to assess backward generalization from modern diffusion/FM generators to legacy TTS systems. This bidirectional setup allows us to rigorously test the synthesis-invariance of our proposed approach. \newline
\noindent\textbf{Training Details}:
All models are trained using the Adam optimizer with cross-entropy loss. The learning rate is set to 1e-3, with a batch size of 32 for 50 epochs. Dropout and early stopping are applied to prevent overfitting. We use five-fold cross-validation within each dataset and report average metrics. For cross-dataset evaluation, no target data is used during training.
\begin{table}[!h]
\setlength{\tabcolsep}{4pt}
\scriptsize
\centering
\begin{tabular}{l|c|c|c|c}
\toprule
\multirow{2}{*}{\textbf{PTMs}} & \multicolumn{2}{c|}{\textbf{Baseline}} & \multicolumn{2}{c}{\textbf{\texttt{RHYME}}} \\
\cmidrule(lr){2-3} \cmidrule(lr){4-5}
& \textbf{TR-A TE-D} & \textbf{TR-D TE-A} & \textbf{TR-A TE-D} & \textbf{TR-D TE-A} \\
\midrule
Xvector         & 33.27  & 22.71  & 28.09  & 16.96  \\
WavLM           & 29.37  & 20.05  & \cellcolor{emerald5}21.87  & 13.17  \\
HuBERT          & 33.52  & 25.56  & 28.66  & 16.63  \\
Whisper         & 30.36  & 24.38  & 27.54  & 17.45  \\
Wav2Vec         & \cellcolor{emerald5}28.42  & \cellcolor{emerald5}19.89  & 22.81  & \cellcolor{emerald5}12.46  \\
PaSST           & \cellcolor{emerald10}27.24  & \cellcolor{emerald10}17.23  & \cellcolor{emerald10}20.38  & \cellcolor{emerald10}11.59  \\
USAD            & \cellcolor{emerald15}\textbf{20.51}  & \cellcolor{emerald15}\textbf{15.19}  & \cellcolor{emerald15}\textbf{14.12}  & \cellcolor{emerald15}\textbf{10.26}  \\
\bottomrule
\end{tabular}
\caption{EER (\%) comparison of pretrained models (PTMs) under mismatched training-testing conditions. TR-A TE-D: Train on ASVP, Test on DFADD; TR-D TE-A: Train on DFADD, Test on ASVP. Lower EER indicates better performance.}
\label{tab-1}
\end{table}
\begin{table*}[!h]
\setlength{\tabcolsep}{10.5pt}
\scriptsize
\centering
\begin{tabular}{l|c|c|c|c|c|c|c}
\toprule
\multirow{2}{*}{\textbf{Synthesizer}} & \multicolumn{7}{c}{\textbf{\texttt{RHYME} USAD}} \\
\cmidrule(lr){2-8}
                                      & \textbf{ASVP} & \textbf{DFADD} & \textbf{D1} & \textbf{D2} & \textbf{D3} & \textbf{F1} & \textbf{F2} \\
\midrule
\textbf{VoiceBox} \cite{le2023voicebox}        & 31.32 & 10.38    & 48.17 & 56.83 & 44.91 & 24.26 & 31.45 \\
\textbf{VoiceFlow} \cite{guo2024voiceflow}       & 24.78 & 4.02    & 29.62 & 27.39 & 28.55 & 38.04 & 10.93 \\
\textbf{NaturalSpeech3} \cite{ju2024naturalspeech}  & 11.63 & 1.79   & 27.45 & 50.07 & 57.89 & 10.38 & 8.56  \\
\textbf{Causal Multi-scale TTS} \cite{li2024cm}           & 20.92 & 0   & 14.14 & 1.09  & 0.69  & 0.019 & 0.02  \\
\textbf{DiffProsody} \cite{oh2024diffprosody}     & 19.18 & 3.46    & 53.77 & 20.85 & 17.66 & 14.51 & 13.24 \\
\textbf{Diffar} \cite{benita2023diffar}          & 32.47 & 0   & 67.38 & 48.22 & 19.84 & 13.79 & 0.13  \\
\textbf{DiTTo-TTS} \cite{lee2024ditto}        & 10.34 & 0.37    & 24.51 & 23.07 & 14.88 & 16.92 & 5.37  \\
\textbf{ReFlow-TTS} \cite{guan2024reflow}      & 13.69 & 0    & 29.02 & 26.85 & 16.43 & 15.27 & 3.08  \\
\bottomrule
\end{tabular}
\caption{Performance on \textbf{\texttt{RHYME}} framework using the USAD backbone, evaluated under the proposed unseen-synthesis generalization protocol. Each row corresponds to a different speech synthesizer, while columns represent test domains: ASVspoof, DFADD, and five held-out subsets (D1–D3: diffusion-based; F1–F2: flow-matching based). All values are reported as Equal Error Rate (EER \%), where lower scores indicate better generalization and spoof detection capability.}
\label{tab-2}
\end{table*}

\subsection{Experimental Results}
\begin{figure}[!h]
    \centering
    \subfloat[]{%
        \includegraphics[width=0.23\textwidth]{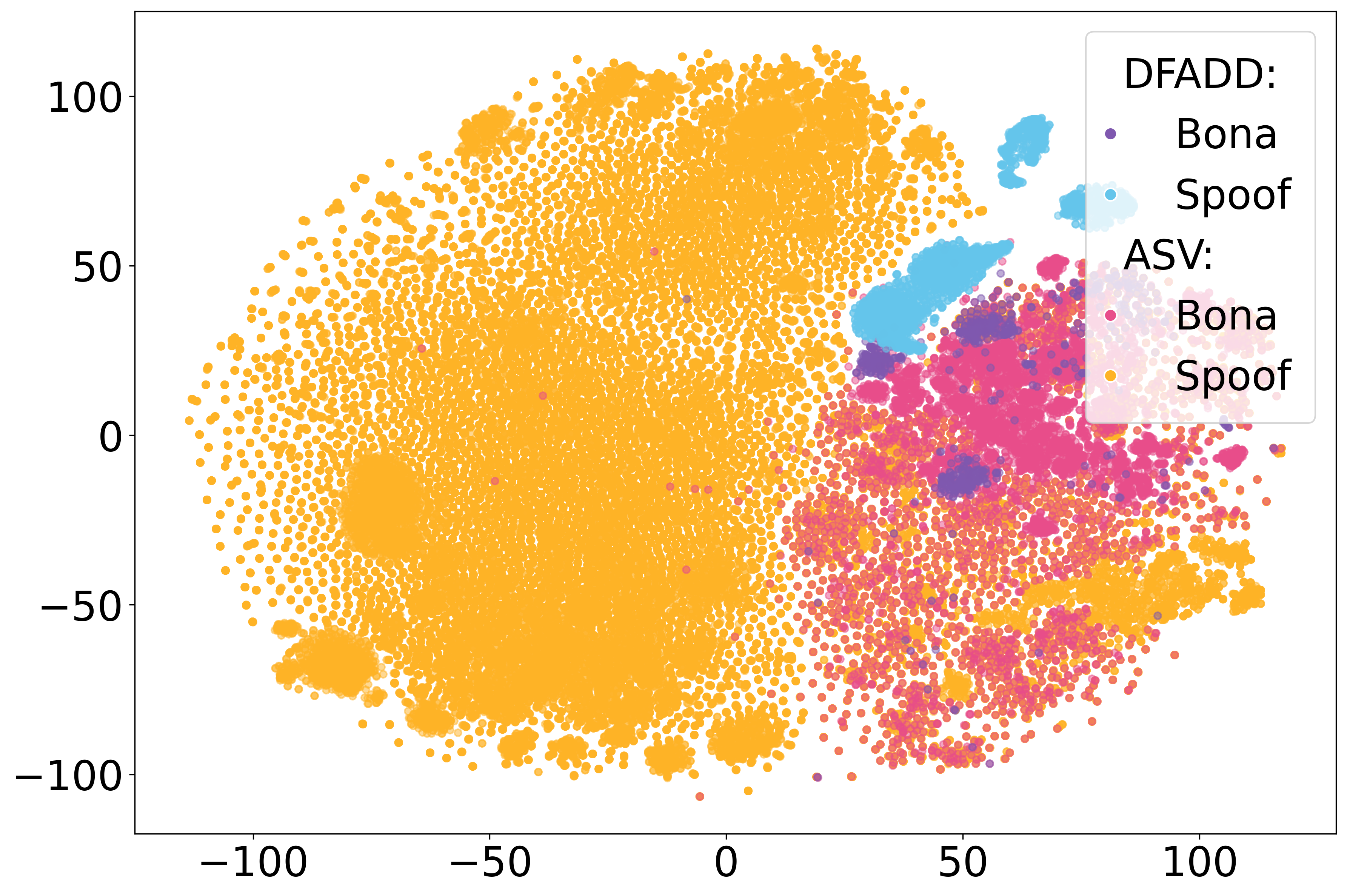}
        \label{fig:tsne_a}
    }
    \subfloat[]{%
        \includegraphics[width=0.23\textwidth]{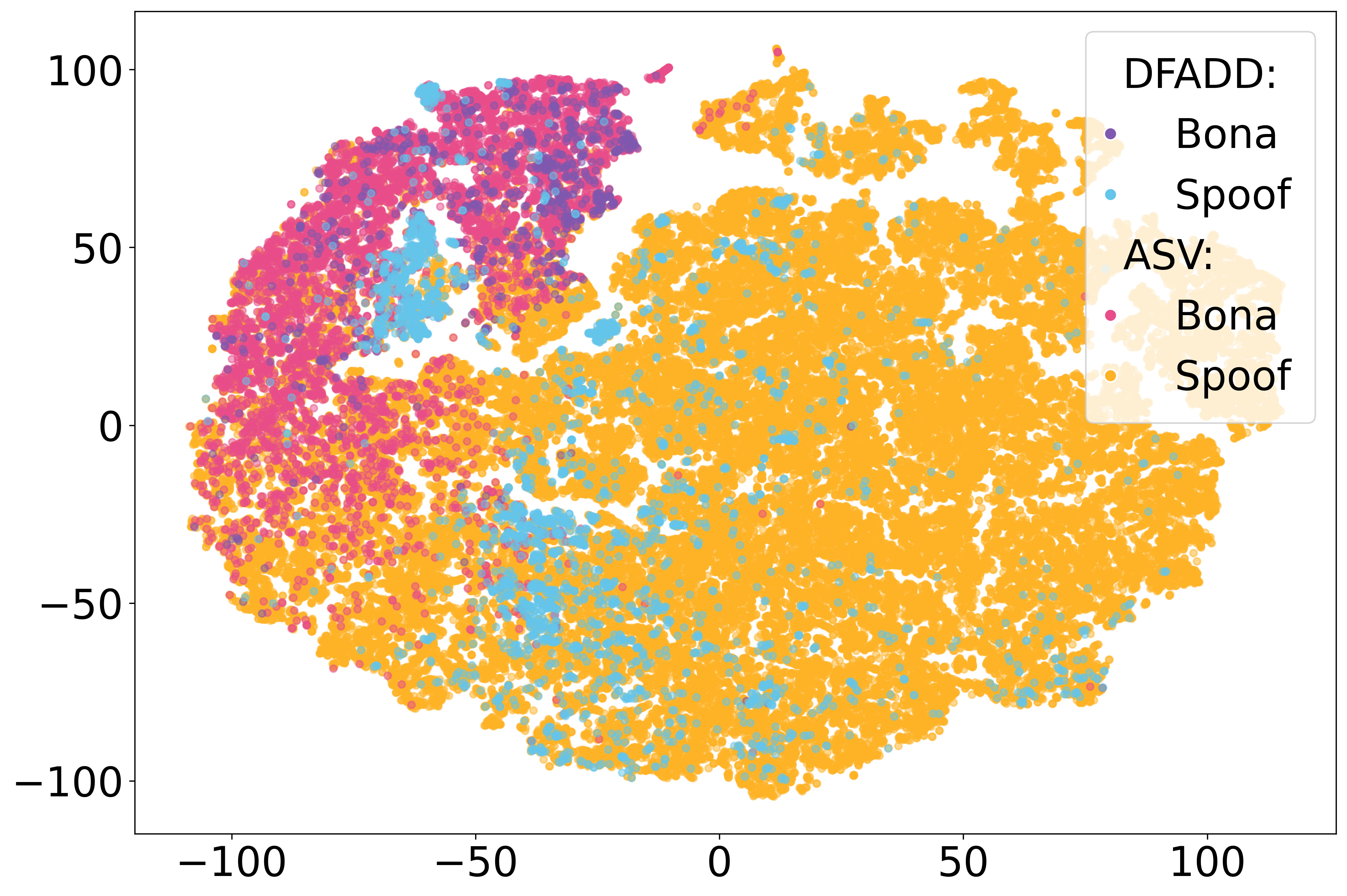}
        \label{fig:tsne_b}
    }
   \caption{t-SNE of (a) Raw USAD embeddings on DFADD+ASV, (b)  fused embeddings (RHYME) trained on DFADD  and tested on DFADD and ASVspoof; colors indicate real vs. fake labels\texttt{RHYME}}
    \label{fig:2}
\end{figure}

\begin{figure}[!hbt]
    \centering

    \subfloat[]{%
        \includegraphics[width=0.23\textwidth]{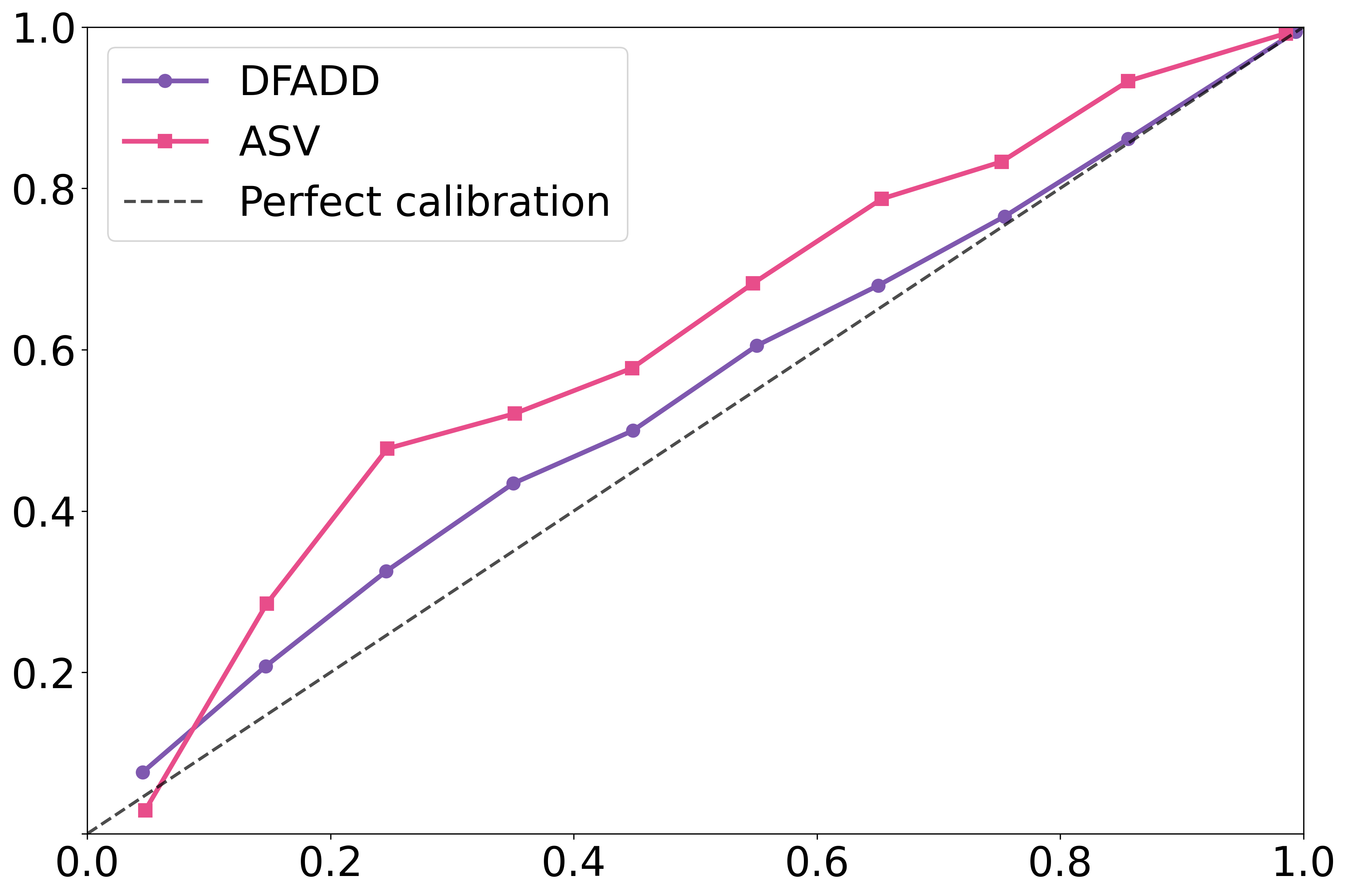}
        \label{fig:graph_a}
    }
    \subfloat[]{%
        \includegraphics[width=0.23\textwidth]{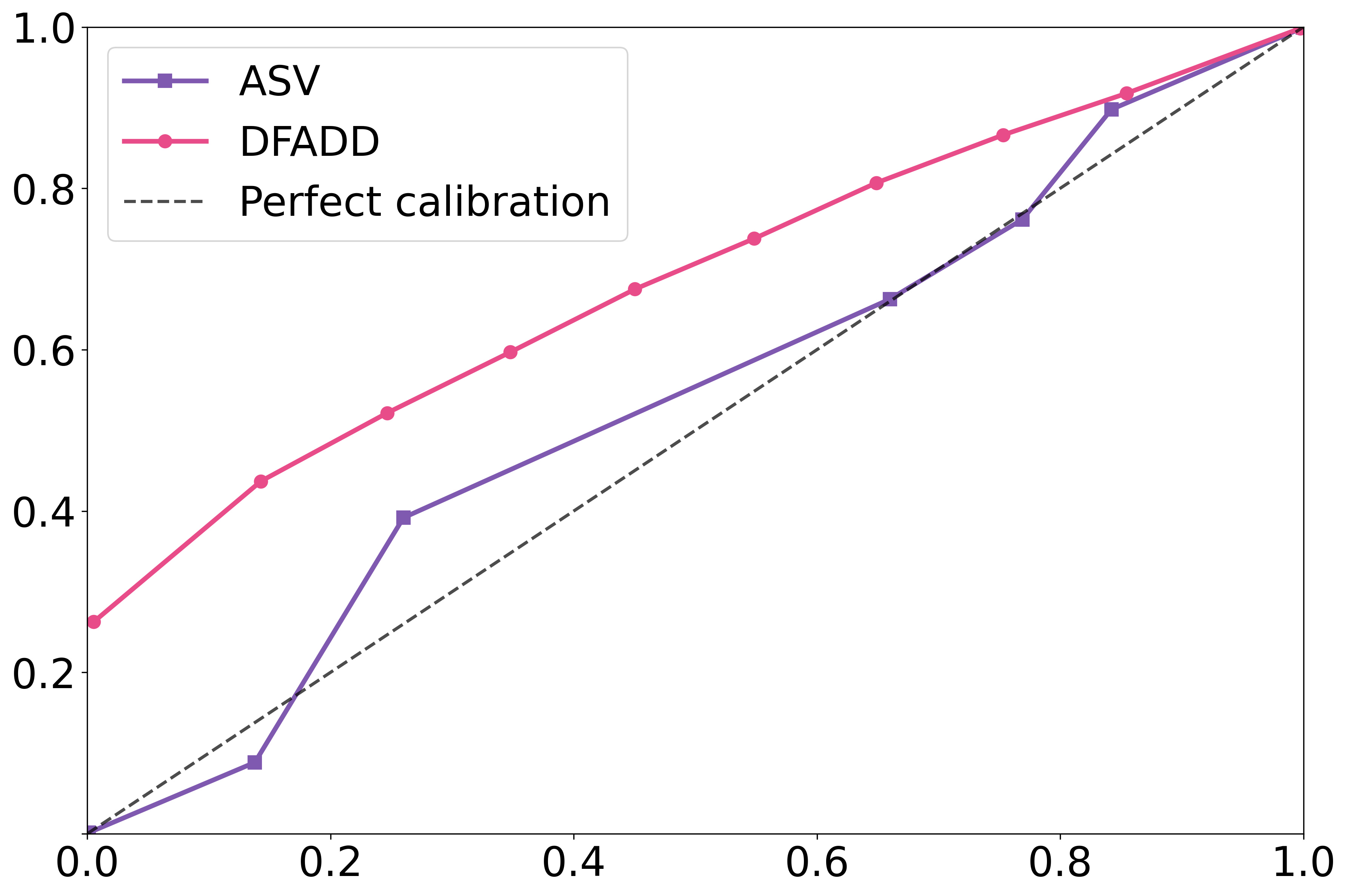}
        \label{fig:graph_b}
    }
   \caption{(a) Trained on DFADD, then tested on both DFADD and ASV test splits. (b) Trained on ASVspoof, then tested on both ASVspoof and DFADD test splits.}
    \label{fig:3}
\end{figure}

Table~\ref{tab-1} shows the EER (\%) results for each model across two cross-domain setups: training on ASVspoof and testing on DFADD (TR-A → TE-D), and the reverse (TR-D → TE-A). Each PTM is assessed in both its original form (Baseline) and its \textbf{\texttt{RHYME}}-enhanced version (Novel). From the results, we observe that across both settings, the RHYME-enhanced models consistently achieve lower EERs than their baseline counterparts. For instance, WavLM benefits from a noticeable EER drop from 29.37\% to 21.87\% (TR-A → TE-D), and from 20.05\% to 13.17\% (TR-D → TE-A). Similarly, models like HuBERT, Whisper, and Wav2Vec2 also show clear improvements under both configurations. These trends confirm that RHYME effectively helps in aligning diverse cues, leading to better generalization under unseen conditions. A consistent observation is that training on DFADD and testing on ASVspoof (TR-D → TE-A) results in relatively lower EERs than the reverse direction. This suggests that the diverse spoofing methods and higher-quality generation found in DFADD lead to more transferable representations. Among all models, USAD achieves the strongest results with EERs of 14.12\% (TR-A → TE-D) and 10.26\% (TR-D → TE-A), showcasing RHYME's ability to handle significant domain shifts. It is also important to note that the DFADD paper \cite{du2024dfadd} reports an average 32.44\% EER for the state-of-the-art end-to-end model AASIST-L under the TR-A → TE-D setup. While our framework does not perform full model fine-tuning and instead relies on frozen pretrained embeddings, our framework achieves a substantially lower EER of 14.12\%—underscoring the efficacy of geometry-aware fusion for robust and synthesis-invariant detection. \newline
\noindent In Table~\ref{tab-2}, we evaluate the performance of \textbf{\texttt{RHYME}}, which is specifically designed to assess generalization in truly unseen conditions.  In this setup, we first extract embeddings using USAD backbone PTMs from both ASVspoof and DFADD datasets. These embeddings are used to train downstream models using our \textbf{\texttt{RHYME}} framework. The testing phase involves audio samples synthesized from various TTS models—which were not part of the training data. These samples were collected from official demo pages of each model, introducing a strong domain shift due to differences in speaker identity, acoustic environments, and generation fidelity. Since the model was only trained on the curated subsets from DFADD, it did not encounter any of the demo-sourced speech samples during training, making this a challenging and realistic generalization test. Despite the difficulty of this setting, \textbf{\texttt{RHYME}} achieves strong performance across several domains. Among the generators, CMTTS consistently shows the lowest EERs, including 0\% on DFADD, 14.14\% on D1, and as low as 0.019\% on F1. DiTTo-TTS and ReFlow-TTS also perform well across most domains, indicating RHYME’s ability to handle both autoregressive and diffusion-based synthesis techniques. On comparatively more variable systems like Diffr and DiPro, we observe a wider spread in EERs across domains; however, \textbf{\texttt{RHYME}} still maintains stability, with EERs well below 20\% in most cases. To further understand our framework robustness, we conduct an additional set of experiments where the model is trained on data generated from a single synthesizer—such as D1, D2, D3, F1, or F2—and tested on unseen samples from other models like Voicebox and NaturalSpeech3. This controlled setup isolates how well the model can transfer from one generator’s characteristics to another. Even with such limited training data, RHYME generalizes effectively in many cases, suggesting that the learned representations are transferable across diverse generation styles. We further investigate the representational quality and decision reliability of RHYME using t-SNE visualizations and calibration plots. Figure~\ref{fig:2} illustrates the distribution of embeddings using t-SNE. In subfigure \ref{fig:tsne_a}, we visualize raw USAD embeddings extracted from DFADD and ASVspoof. The overlap between real and fake samples suggests weak discrimination in the original embedding space. While, subfigure \ref{fig:tsne_b} shows RHYME-fused embeddings when trained on DFADD and evaluated on both DFADD and ASVspoof. Here, the real and fake classes form visibly distinct clusters, highlighting RHYME’s ability to disentangle domain-invariant and spoof-discriminative cues through geometric fusion. To evaluate how well model confidence aligns with actual correctness, we use calibration diagrams shown in Figure~\ref{fig:3}. Subfigure \ref{fig:graph_a} depicts the calibration curve when the model is trained on DFADD and tested on both DFADD and ASVspoof. The curve closely follows the diagonal, indicating good calibration between predicted probabilities and true outcome frequencies, more importantly, the reliability curve remains closely aligned for the out-of-domain ASVspoof test split. This demonstrates that the model’s softmax outputs remain trustworthy even under out-of-domain detection. On the other hand, subfigure \ref{fig:graph_b}, where the model is trained on ASVspoof, shows larger deviations—highlighting miscalibration. These results reinforce that RHYME not only improves detection accuracy but also produces more reliable probability estimates, a key requirement for deployment in risk-sensitive applications.  \par

\subsection{Evaluation Setup}
We evaluate the effectiveness and generalization ability of \textbf{\texttt{RHYME}} under two settings, we use multiple pretrained models (PTMs). We extract embeddings from these PTMs and train models on dataset (ASVspoof or DFADD), then test on the other. This helps us understand how well each model adapts to changes in data distribution and spoofing style. In the second setup, we test the model on deepfakes generated by a variety of TTS systems not seen during training. These samples are collected from the official demo pages of VoiceBox \cite{le2023voicebox}, VoiceFlow \cite{guo2024voiceflow}, NaturalSpeech 3 \cite{ju2024naturalspeech}, CMTTS \cite{li2024cm}, DiffProsody \cite{oh2024diffprosody}, DiffAR \cite{benita2023diffar}, DiTTo-TTS \cite{lee2024ditto}, and ReFlow-TTS \cite{guan2024reflow}. VoiceBox and VoiceFlow use flow-matching, while the others are diffusion-based. The model was trained only on DFADD and ASVspoof data, so this setup gives us a clear view of how well it performs in truly unseen scenarios. We also perform a focused experiment where the model is trained on samples (D1, D2, D3, F1, F2) and then tested on other synthesizers. This helps isolate how well proposed framework generalizes across different types of generators.

\subsection{Ablation Study}
To quantify the impact of individual components within the proposed \textbf{\texttt{RHYME}} framework, we perform a detailed ablation study along three dimensions: architectural geometry (spherical and hyperbolic branches), fusion strategy (with and without geometry), and integration mechanism (gating). Results are reported in Table~\ref{tab:ablation} under two cross-domain setups: training on ASVspoof and testing on DFADD (TR-A~$\rightarrow$~TE-D), and vice versa (TR-D~$\rightarrow$~TE-A). The full \textbf{\texttt{RHYME}} configuration achieves the best performance, with EERs of 14.12\% and 10.26\%, clearly outperforming all ablated variants. When the geometry-aware branches are removed, performance degrades—removing the spherical branch increases EER to 18.01\% (TR-A~$\rightarrow$~TE-D), and removing the hyperbolic branch results in 17.88\%. This confirms that both geometric projections contribute complementary and discriminative information for detecting synthetic speech. Replacing the RHYME fusion mechanism with a naive Euclidean fusion further degrades performance (19.33\%, 15.12\%), indicating that conventional feature concatenation fails to capture the curved manifold structure inherent in speech embeddings across domains. We also examine the role of the gated fusion mechanism. Disabling it by fixing the gating scalar to $\alpha = 0.5$ leads to higher EERs (19.94\%, 14.78\%), suggesting that the learned weighting between the two branches is crucial for adaptively balancing domain cues.
Finally, using a single PTM (USAD only) without any fusion results in the weakest performance (20.51\%, 15.19\%), reinforcing the importance of \textbf{\texttt{RHYME}} multi-geometry, multi-PTM architecture. Together, these results highlight that each component of \textbf{\texttt{RHYME}} contributes meaningfully, and that their combination yields a substantial gain in generalization under mismatched and unseen conditions.

\noindent
The observed non-linear improvement when both geometric branches are combined stems from their complementary roles rather than redundancy. 
The hyperbolic branch captures hierarchical generator relationships, while the spherical branch encodes periodic and energy-invariant synthesis artifacts; these cues occupy largely orthogonal representational subspaces. 
When fused through Riemannian barycentric averaging, the interaction between curvature-aware embeddings yields a super-additive gain—greater than the sum of their individual effects. 
We confirmed this consistency across five random seeds, reporting mean~$\pm$~standard-deviation values in Table~\ref{tab:ablation} to ensure statistical reliability.

\begin{table}[!ht]
\centering
\setlength{\tabcolsep}{6pt}
\scriptsize
\begin{tabular}{@{} l|c|c @{}}
\toprule
\textbf{Configuration} & \textbf{TR-A $\rightarrow$ TE-D} & \textbf{TR-D $\rightarrow$ TE-A} \\
\midrule
\textbf{\texttt{RHYME}}                            & \textbf{14.12} & \textbf{10.26} \\
No Gating ($\alpha = 0.5$)           & 19.94          & 14.78          \\
No Spherical Branch                  & 18.01          & 14.22          \\
No Hyperbolic Branch                 & 17.88          & 13.89          \\
Euclidean Fusion (No Geometry)      & 19.33          & 15.12          \\
Baseline (USAD only)              & 20.51          & 15.19          \\
\bottomrule
\end{tabular}
\caption{Equal Error Rate (EER \%) evaluation under distribution shift. We assess the effect of removing RHYME components: gating, spherical and hyperbolic branches, and geometry-aware fusion. Results demonstrate that each module contributes to performance, with the \textbf{\texttt{RHYME}} configuration achieving the lowest EERs.}
\label{tab:ablation}
\end{table}

\section{Conclusion}
In this work, we presented \textbf{\texttt{RHYME}}, a unified and geometry-aware framework for generalizable audio deepfake detection across diverse synthesis paradigms, including TTS, vocoder, and diffusion-based generators. By leveraging hyperbolic and spherical projections to model complementary synthesis artifacts and fusing them via Riemannian barycentric averaging, the proposed method learns a synthesis-invariant embedding space that supports strong generalization. The outcomes confirm that uniting hyperbolic hierarchy modeling with spherical artifact encoding through Riemannian fusion yields a geometry-driven, synthesis-invariant detector that generalizes across unseen generators. Extensive experiments show that it consistently outperforms existing detectors and fusion baselines under cross-corpus, zero-shot, and unseen-generator settings. These findings demonstrate the effectiveness of our approach and highlight its potential as a reliable and scalable solution for audio deepfake detection under diverse and previously unseen conditions. Our work also calls upon researchers to build on our extended benchmarks to further advance performance in generalizable audio deepfake detection.

\section*{Limitations}
Our framework achieves strong generalization across diverse synthesis paradigms—including TTS, vocoder, and modern diffusion or flow-matching generators—but there are certain limitations to our study. First, our evaluation is restricted to English-language datasets; generalization to multilingual and accented speech, including cross-lingual and code-switched scenarios, remains unexplored.

\section*{Broader Impact and Ethics}
This work aims to support the growing need for detecting synthetic speech by introducing a generalizable deepfake detection framework. As audio generation models become more realistic, tools like \textbf{\texttt{RHYME}} can help protect against misuse in areas such as voice fraud, impersonation, and misinformation. At the same time, we recognize the importance of using such detection systems responsibly. Our current study is limited to English datasets, and we encourage future research to consider more diverse languages and speaker populations. We also acknowledge that deepfake detection technology could be misused, for example, in surveillance or censorship. To reduce such risks, our work is shared for research purposes only and does not involve any sensitive or private data. We believe this study provides a step forward in building safer and more trustworthy speech systems.

\bibliography{custom}

\end{document}